\documentclass[a4paper,12pt]{article}
\usepackage{epsfig}
\title{\Large\bf A TIME VARYING STRONG COUPLING CONSTANT AS A MODEL OF
INFLATIONARY UNIVERSE}

\author
{ \it \bf  N. Chamoun$^{1,2}$, S. J. Landau$^{3,4}$\thanks{fellow of
CONICET} , H. Vucetich$^{1,3}$\thanks{member of CONICET} ,
\\ \small$^1$ Departamento de F\'{\i}sica, Universidad Nacional de
La Plata,\\ \small c.c. 67, 1900 La Plata, Argentina. \\
\small$^2$ Department of Physics, Higher Institute for Applied
Sciences and Technology,\\ \small P.O. Box 31983, Damascus, Syria.
\\ \small$^3$ Observatorio Astron\'{o}mico, Universidad Nacional
de La Plata,\\ \small Paseo del Bosque S$/$N, CP 1900 La Plata,
Argentina.
\\ \small$^4$ Departamento de F\'{\i}sica, Universidad Nacional de
Buenos Aires,\\ \small  Ciudad Universitaria - Pab. 1, 1428 Buenos Aires,
  Argentina \\
}

\date{}

\begin{document}
\maketitle
\begin{abstract}
We consider a scenario where the strong coupling constant was
changing in the early universe. We attribute this change to a
variation in the colour charge within a Bekenstein-like model.
Treating the vacuum gluon condensate $<G^2>$ as a free parameter ,
we could generate inflation, similar to a chaotic inflation
scenario, with the required properties to solve the fluctuation
and other standard cosmology problems. A possible approach to end
the inflation is suggested.
\end{abstract}

{\bf Key words}: fundamental constants; inflationary models; cosmology; QCD

{\bf PACS}: 98.80Cq,98.80-k,12.38-t

\section{Introduction}
\label{Introduction} The hypothesis that the universe underwent a
period of exponential expansion at very early times is the most
popular theory of the early universe. The ``old" \cite{guth81} and
``new" \cite{linde82,albrecht82} inflationary universe models were
able to solve the ``horizon", ``flatness" and ``structure
formation" problems creating, in turn, their own problems, some of
which the ``chaotic" model \cite{linde83} and its extensions
\cite{linde94} could tackle. On the other hand, these kind of
cosmological models are favored by recent measurements from the
Cosmic Microwave Background (CMB) \cite{deBernardis,MAXIMA} and
the WMAP results \cite{wmap}. Usually, inflationary models are all
based on the use of new fundamental scalar fields, the
``inflatons", which can not be the Higgs fields of ordinary gauge
theories. Later, other possible alternatives to inflationary
cosmology were proposed. Rather than changing the matter content
of the universe, new concepts were adopted such as non-commutative
space-time quantizations \cite{alexander01}, brane settings of a
cycling universe \cite{khoury}, and a change in the speed of light
in the early universe \cite{moffat93,albrecht99} or, more
generally, a varying fine structure constant $\alpha_{em}$
\cite{barrow98}. In \cite{chamoun2000}, we have generalized
Bekenstein model \cite{bekenstein82} for the time variation of
$\alpha_{em}$ to QCD strong coupling constant $\alpha_S$ and found
that experimental constraints going backward till quasar formation
times rule out $\alpha_S$ variability. In this letter, we discuss
how our ``minimal" Bekenstein-like model for $\alpha_S$, when
implemented in the very early universe, can provide a realization
of inflation driven by the trace anomaly of QCD energy momentum
tensor. Inflationary models driven by the trace anomaly of
conformally coupled matter fields are treated in the literature
\cite{hawking00} while in our toy model we concentrate on the
gauge fields of QCD. Assuming the universe is radiation-dominated
at early times and that vacuum expectation values predominate over
matter densities, we find, with suitable values of the gluon
condensate $<G^2>$ far larger than its present value and the
Bekenstein length scale $l$ smaller than the Planck-Wheeler length
scale $L_P$, that our model for inflation is self-consistent with
acceptable numerical results to solve the fluctuation and other
problems. We find that the QCD lagrangian with a changing strong
coupling constant leads naturally to a monomial quadratic
potential like the chaotic scenario. However, while the large
values of the inflaton matter field plague the latter scenario,
they just amount in our model to a reduction of the strong charge
by around $10$ times during the inflation. We shall not dwell on
the possible mechanisms by which the gluon condensate could have
decreased to reach its present value, and which might be necessary
in order not to have exotic relics. Rather we wish to concentrate
on the conditions we should impose on our model to be interesting.
We hope this phenomenological approach could prompt further work
on a possible connection between time-varying fundamental
``constants"and inflationary models.

\section{Analysis}
\label{analysis} We follow the notations of \cite{chamoun2000}.
Therein, we used the QCD Lagrangian with a varying coupling
``constant"
\begin{eqnarray}
L_{QCD}&=& L_{\epsilon}+L_g+L_m \nonumber\\
&=&-\frac{1}{2l^2}\frac{\epsilon_{,\mu}
\epsilon^{,\mu}}{\epsilon^2}
 -\frac{1}{2}Tr(G^{\mu\nu}G_{\mu\nu}) +
\sum_f \bar{\psi}^{(f)}(i \gamma^\mu D_\mu -M_f) \psi^{(f)}
\end{eqnarray}
where the metric used for raising and lowering indices is the R-W
metric, $l$ is the Bekenstein scale length, $\epsilon(x)$ is a
scalar gauge-invariant and dimensionless field representing the
variation of the strong coupling ``constant"
$g(x)=g_0\epsilon(x)$, $D_{\mu} =
\partial_{\mu} - i g_0 \epsilon(x) A_{\mu}$ is the covariant
derivative and where the gluon tensor field is given by:
\begin{eqnarray}
\label{gluon equation} G^a_{\mu\nu} &=& \frac{1}{\epsilon}
[\partial_\mu(\epsilon A^a_\nu)-\partial_\nu(\epsilon
A^a_\mu)+g_0\epsilon^2f^{abc}A^b_\mu A^c_\nu]
\end{eqnarray}
We assume homogeneity and isotropy for an expanding universe and
so consider only temporal variations for $\alpha_S\equiv
\frac{g^2(t)}{4\pi}=\alpha_{S_0}\epsilon^2(t)$. We assume also,
rather plausibly in the radiation-dominated early universe,
negligence of matter contribution to get the following equations
of motion
\begin{eqnarray}
\label{equation1}
(\frac{G^{\mu\nu}_a}{\epsilon})_{;\mu}-g_0f^{abc}G^{\mu\nu}_b
A^c_\mu+\sum_f g_0\bar{\psi}t^a \gamma^\nu\psi &=&0
\end{eqnarray}
\begin{eqnarray}
\label{equation2} (a^3 \frac{\dot{\epsilon}}{\epsilon})^.
&=&\frac{a^3(t) l^2}{2} \langle G^2 \rangle
\end{eqnarray}
where $a(t)$ is the expansion scale factor in the R-W metric.

Computing the canonical energy-momentum tensor $\frac{\partial
L}{\partial(\partial_{\alpha}\phi_i)}\partial^{\beta}\phi_i -
g^{\alpha\beta}L$ we get a non gauge invariant quantity. This may
be cured by a standard technique \cite{itzykson80} amounting to a
subtraction of a total derivative and hence not changing the
equations of motion. We subtract the total derivative $ \Delta
T^{\alpha\beta}=\partial_{\nu}\left(
-\frac{G_a^{\alpha\nu}}{\epsilon}\epsilon A^{a \beta}\right) $ to
get, with the use of equation (\ref{equation1}), the
gauge-invariant energy momentum tensor
\begin{eqnarray}
T^{\alpha\beta}&=& G_a^{\alpha\nu}G_{\nu}^{a\beta}+ i\sum_f
\bar{\psi}^{(f)} \gamma^{(\alpha} D^{\beta
)} \psi^{(f)}
-
\frac{1}{l^2}\frac{\partial^{\alpha}\epsilon
\partial^{\beta}\epsilon}{\epsilon^2} \nonumber\\&&
-g^{\alpha\beta} \left[ -\frac{1}{4}G^{\mu\nu}_a G_{\mu\nu}^a +
\sum_f \bar{\psi}^{(f)}(i \gamma^\mu D_\mu -M_f) \psi^{(f)}
-\frac{1}{2l^2}\frac{\epsilon_{,\mu}\epsilon^{,\mu}}{\epsilon^2}
 \right]
\end{eqnarray}

Since we assume radiation dominated era during the very early
universe we can concentrate on the gauge fields and neglect the
matter fields contribution and so we decompose our energy momentum
tensor into two parts: the gauge part and the $\epsilon$-scalar
field part
\begin{eqnarray}
T_{\alpha\beta}&=&T_{\alpha\beta}^{g}+T_{\alpha\beta}^{\epsilon}
\\&=& \left ( G_{\alpha\nu}G^{\nu}_{\beta}-g_{\alpha\beta}
\left[ -\frac{1}{4}G^{\mu\nu}_a G_{\mu\nu}^a \right] \right) +
\left(
-
\frac{1}{l^2}\frac{\partial_{\alpha}\epsilon
\partial_{\beta}\epsilon}{\epsilon^2}-g_{\alpha\beta} \left[
-\frac{1}{2l^2}\frac{\epsilon_{,\mu}\epsilon^{,\mu}}{\epsilon^2}
\right] \right)\nonumber
\end{eqnarray}
Here all the operators are supposed to be renormalized and it is
essential in the inflationary paradigm that quantum effects are
small in order to get small fluctuations in the CMB, which will be
seen to be the case in the model.

The contribution of the scalar field to the energy density
$\rho_{\epsilon}=T_{00}^{\epsilon}$ and to the pressure
$T^{\epsilon}_{ij}=g^{(3)}_{ij}p_{\epsilon}$ can be computed and
we get
\begin{eqnarray}
\rho_{\epsilon}&=-\frac{1}{2l^2}(\frac{\dot{\epsilon}}{\epsilon})^2=&p_{\epsilon}
\end{eqnarray}

On the other hand, the gauge field contribution
$T_{\alpha\beta}^{g}$ can be decomposed into trace and traceless
parts. In such a way, we can write the corresponding energy mass
density $\rho_{g}=T^{g}_{00}$ and the pressure $p_{g}$, like in
``ordinary" QCD, as a sum of two terms:

\begin{eqnarray}
\rho_{g} &=& \rho^r_{g} + \rho^{T}_{g} \\
p_{g} &=& p^r_{g} + p^T_{g}
\end{eqnarray}
where $\rho_{g}^{r}$ is the density corresponding to the
``traceless" part of the gauge field satisfying
\begin{eqnarray}
\rho_{g}^{r} & = & 3  p_{g}^{r}
\end{eqnarray}
while the trace part of the gauge field energy-momentum tensor is
proportional  to $g_{\alpha \beta}$ and behaves like a
`cosmological constant' term. Thus, the corresponding equation of
state reads:
\begin{eqnarray}
\label{impose} \rho^{T}_{g} &=& -p^{T}_{g}
\end{eqnarray}
We shall assume, here, that the trace anomaly relation for $T^{\mu
g}_\mu$ is the same as the ``ordinary'' QCD trace anomaly. This
can be justified because the energy-momentum tensor $T^{g}_{\alpha
\beta}$ is identical in form to ``ordinary" QCD and that the trace
anomaly which involves only gauge invariant quantities should, by
dimensional analysis, be proportional to $G^2$. However, we have
checked that a change in the numerical value of the
proportionality factor would not alter our conclusions. Thus we
take, up to leading order in the time-varying coupling ``constant"
$\alpha_S=\alpha_{S_0}\epsilon^2$, the relation \cite{greiner95}:
\begin{eqnarray}
\label{traceanomaly}
 T^{\mu g}_\mu &=&
\rho_{g}-3p_{g} = -\frac{9 \alpha_{S_0} \epsilon^2}{8 \pi}
G^{\mu\nu}_a G^a_{\mu\nu}
\end{eqnarray}
Again, neglecting matter contribution during the radiation
dominated era, we limit our gluon operator $G^{\mu\nu}_a
G^a_{\mu\nu}$ matrix elements to its condensate vacuum expectation
value $<G^2>$, and so we get:
\begin{eqnarray}
\label{rhoT} \rho^{T}_{g}&=&-\frac{9 \alpha_{S_0} }{32 \pi}
\epsilon^2 <G^2>
\end{eqnarray}
Hence, we can write the total energy mass density as
\begin{eqnarray}
\rho = \rho_{g}^{r}+ \rho_{g}^{T}+ \rho_{\epsilon}
\end{eqnarray}
and equation (\ref{impose}) would suggest, in analogy to ordinary
inflationary models, that the QCD trace anomaly could generate the
inflation. For this, let us assume that the ``trace-anomaly"
energy mass density contribution is much larger than the other
densities:
\begin{eqnarray}
\rho_{g}^{T} >> \rho_{\epsilon},\rho^{r}_{g} &\Rightarrow& \rho
\sim \rho_{g}^{T}
\end{eqnarray}
Then, equation (\ref{rhoT}) tells that the vacuum gluon condensate
$<G^2>$ should have a negative value which is not unreasonable
since the inflationary vacuum has ``strange" properties. In
ordinary inflationary models, it is filled with repulsive-gravity
matter turning gravity on its head \cite{guth01}. This reversal of
the vacuum properties is reflected, in our model, by a reversal of
sign for the vacuum gluon condensate.

Now we seek a consistent inflationary solution  to the FRW
equations in a flat space-time:
\begin{eqnarray}
\label{frw}
(\frac{\dot{a}}{a})^2&=& \frac{8 \pi G_N}{3} \rho\\
\frac{\ddot{{a}}}{a}&=&-\frac{4 \pi G_N}{3}(\rho+3 p)
\end{eqnarray}
where $G_N$ is Newton's constant. The first FRW equation with
(\ref{rhoT}) will give

\begin{eqnarray}
\label{H}
H \equiv \frac{\dot{a}}{a} &=& \epsilon\sqrt{\frac{3 \alpha_{S_0} }{4} G_N |<G^2>|}
\end{eqnarray}
On the other hand, the equation of motion (eq.\ref{equation2}) of the scalar field  can be expressed in the following way:
\begin{eqnarray}
\label{epsilon}
\frac{\ddot{\epsilon}}{\epsilon} + 3 H \frac{\dot\epsilon}{\epsilon}
-(\frac{\dot{\epsilon}}{\epsilon})^2 = \frac{l^2 <G^2>}{2}
\end{eqnarray}
This equation differs from the ordinary `matter' inflationary
scenarios in the term $(\frac{\dot{\epsilon}}{\epsilon})^2$.
However, for ``slow roll'' solutions we neglect the terms
involving $\frac{\ddot\epsilon}{\epsilon}$ and
$(\frac{\dot\epsilon}{\epsilon})^2$ to get
\begin{eqnarray}
\label{potential} 3 H \dot\epsilon &=\frac{l^2 <G^2>}{2}
\epsilon&=-V'(\epsilon)
\end{eqnarray}
which is the same as the ``slow roll'' equation of motion of the
inflaton in ordinary scenarios. In our model, the ``slow roll''
condition can be written as:
\begin{eqnarray}
\label{conscond1} \delta \equiv |\frac{\dot{\epsilon}}{H
\epsilon}| &=\frac{2}{9 \alpha_{S_0}}
(\frac{l}{L_P})^2\frac{1}{\epsilon^2}&<< 1
\end{eqnarray}
This gives us the first hint that Bekenstein hypothesis ($\L_P <
l$) might not survive. Now, we set $\epsilon_f$, the value of
$\epsilon$-field at the end of inflation $t_f$, to $1$ so that the
time evolution of the strong coupling terminates with the end of
inflation. We expect for ``slow roll'' solutions that
$\epsilon_i$, the value of $\epsilon$ at the initial time of
inflation $t_i$ corresponding to when the CMB modes freezed out,
to be of order $1$. Taking this into account, and in order that
the changes of the Hubble constant and the energy mass density are
not very large during the inflation, we assume the gluon
condensate value $<G^2>$ to stay approximately constant during
much of the inflation. Then we get a linear evolution in time for
the $\epsilon$- scalar field
\begin{eqnarray}
\label{epsevo} \epsilon (t) & = & \epsilon_i
-\frac{1}{3^{3/2}(\alpha_{S_0})^{1/2}}(\frac{l}{L_p})^2 G^{1/2}_N
|<G^2>|^{1/2} (t-t_i)
\end{eqnarray}
and we have, as in chaotic scenarios, a simple quadratic
potential:
\begin{eqnarray}
V(\epsilon)&=&\frac{l^2 |<G^2>|}{4}\epsilon^2
\end{eqnarray}

One can make explicit the correspondence between our model with
$\epsilon$-scalar field and the chaotic scenario with an
$\phi$-inflaton matter field. Looking at equation (\ref{rhoT}) and
comparing with $\rho = {\cal V}(\phi)$ in ordinary inflationary
models, we see that the gluon condensate plays a role of a
potential for the ``inflaton"-$\epsilon$ field. Comparing
equations (\ref{H}) and (\ref{potential}) with the corresponding
relations in ordinary inflationary models:
\begin{eqnarray}
 H^2 &=& \frac{8\pi}{3 M_{pl}^2}G_N {\cal V}(\phi) \\
 3 H \dot\phi &=&-{\cal V}'(\phi)
\end{eqnarray}
and remembering that the dimension of the matter scalar field
$\phi$ is one, we have ${\cal V} \propto \frac{V}{l^2}$ and $\phi
\propto \frac{\epsilon}{l}$, and we can find the relations between
$(\epsilon,V(\epsilon))$ and $(\phi,{\cal V}(\phi))$:
\begin{eqnarray}
\phi = \frac{\sqrt{y}}{l}\epsilon &{\rm with}& y=\frac{9 \alpha_{S_0}}{8 \pi} \\
\frac{y}{l^2}V(\epsilon)&={\cal V}(\phi) =& \frac{l^2 |<G^2>|}{4}
\phi^2
\end{eqnarray}

\section{Results and Conclusion}
Now, we check that our model is able to fix the usual problems of
the standard (big bang) cosmology. First, in order to solve the
``horizon" and ``flatness" problems we need an inflation
$\frac{a(t_f)}{a(t_i)}$ of order $10^{28}$ implying an inflation
period $\Delta t = t_f - t_i$ such that
\begin{eqnarray}
\label{hubbletime} H \;\; \Delta t&\sim& 65
\end{eqnarray}
Furthermore, it should satisfy the constraint
\begin{eqnarray}
\label{timecons} 10^{-40} s \leq \Delta t \ll 10^{-10} s
\label{times}
\end{eqnarray}
 so that not to conflict with the explanation of the
baryon number and not to create too large density fluctuations
\cite{hawking82,hawking85}. The bound $10^{-10} s$ corresponds to the
time, after the big bang, when the electroweak symmetry
breaking took place. Presumably, our inflation should have
ended far before this time.
Thus, from equations (\ref{hubbletime}), (\ref{times}) and (\ref{H}) we get the following
bounds on $|<G^2>|$:
\begin{eqnarray}
\label{bound1} 3 \times 10^7 GeV^2 \ll \epsilon |<G^2>|^{1/2} \leq
3 \times 10^{37} GeV^2
\end{eqnarray}
In order to determine $\epsilon_i$, we have $\frac{d \ln a}{d
\epsilon} = \frac{H}{\dot{\epsilon}} \simeq
-\frac{3H^2}{V'(\epsilon)} \simeq -\frac{8 \pi \rho^T_g}{M_{Pl}^2
V'}$ which gives, using equations (\ref{rhoT}) and
(\ref{potential}), the relation:
\begin{eqnarray}
\label{ei} 65&\sim \ln \frac{a(t_f)}{a(t_i)}=&(\frac{L_P}{l})^2
\frac{9 \alpha_{S_0}}{4}(\epsilon_i^2 -1)
\end{eqnarray}
Next, comes the ``formation of structure"
problem and we require the
fractional density fluctuations at the end of
inflation to be of the order $\frac{\delta M}{M}\mid _{t_f} \sim
10^{-5}$ so that quantum fluctuations
in the de Sitter phase of
the inflationary universe form the source of perturbations
providing the seeds for galaxy formation and in order to agree
with the CMB anisotropy limits. Within the relativistic theory of
cosmological perturbations \cite{brandenberger97}, the above
fractional density fluctuations represent (to linear order) a
gauge-invariant quantity and
 satisfy the equation
\begin{eqnarray}
\label{fluctfinal1} \frac{\delta M}{M}\mid _{t_f} &=& \frac{\delta
M}{M}\mid _{t_i}\frac{1}{1+\frac{p}{\rho}}\mid _{t_i}
\end{eqnarray}
where $\delta M$ represent the mass perturbations.

The initial fluctuations are generated quantum mechanically and, at the
linearized level, the equations describing both gravitational and matter
perturbations can be quantized in a consistent way \cite{mukhanov92}. The
time dependence of the mass is reflected in the nontrivial form
of the
solutions to the mode equations and one can compute the expectation
value of field operators such as the power spectrum and get the following
result for the initial mass perturbation \cite{brandenberger97,mukhanov92}
\begin{eqnarray}
\label{fluctinitial} \frac{\delta M}{M}\mid _{t_i}= \frac{{\cal
V}'(\Phi)H}{\rho}&=&\sqrt{y}\frac{V'(\epsilon)H}{l\rho}
\end{eqnarray}
whence
\begin{eqnarray}
\label{fluctfinal2} 10^{-5} \sim \mid \frac{\delta M}{M}\mid
_{t_f} &=& \sqrt{y} \mid \frac{V' H}{l} \mid _{t_i} \frac
{1}{|(\rho+p)|_{t_i}}
\end{eqnarray}
In order to evaluate $(\rho+p)|_{t_i}$ we use the energy
conservation equation:
\begin{eqnarray}
\label{enercon} \dot{\rho}+3 (\rho+p) \frac{\dot{a}}{a} &=&0
\end{eqnarray}
and after substituting $\rho \sim \rho^T_{g}$ we get
\begin{eqnarray}
\label{initial}
|(\rho+p)|_{t_i} &=& \frac{1}{24 \pi} (\frac{l}{L_P})^2 |<G^2>|
\end{eqnarray}
In fact, the energy conservation equation can be used to solve for
$\rho^r_{g}$ and we could check that
\begin{eqnarray}
\rho^r_{g} (\dot{\rho^r_{g}}) \sim \rho_{\epsilon}
(\dot{\rho_{\epsilon}}) &\sim& \delta \times \rho ^T_{g} (\delta
\times \dot{\rho^T_{g}})
\end{eqnarray}
where $\delta \equiv |\frac{\dot{\epsilon}}{H\epsilon}| \sim
\frac{1}{\epsilon^2}(\frac{l}{L_P})^2$ and so, when the ``slow
roll'' condition (\ref{conscond1}) is satisfied, our solution
assuming the predominance of the ``trace-anomaly" energy mass
density is self-consistent. Substituting equation (\ref{initial})
in (\ref{fluctfinal2}) and using equation (\ref{potential}) we get
\begin{eqnarray}
\label{important} \frac{l}{L_P} &\sim& 9 \sqrt{\frac{3\pi}{2}}
\alpha_{S_0} 10^{5} \epsilon^2 |<G^2>|^{\frac{1}{2}} G_N
\end{eqnarray}
Hence, taking $G_N \sim 10^{-38} GeV^{-2}$ we obtain
\begin{eqnarray}
\label{G2ei} \frac{l}{L_P} \sim
\frac{|<G^2>|^{\frac{1}{2}}}{10^{34} GeV^2} \epsilon^2
\end{eqnarray}
and combining this last result with (\ref{bound1}), we get
\begin{eqnarray}
10^{-27} &<<& \frac{1}{\epsilon}\frac{l}{L_p} \leq 10^3
\end{eqnarray}
The ``slow roll'' condition (\ref{conscond1}) is consistent with
the upper bound, while the lower bound restricts $\epsilon_i$ not
to be too large.

On the other hand, it is possible to calculate the spectral index of the primordial power
spectrum for a quadratic potential as follows:
\begin{eqnarray}
n-1=-4\eta &{\rm where}& \eta = \frac{M_{Pl}^2}{8\pi}\frac{{\cal
V}''}{{\cal V}}=\frac{2}{9 \alpha_{S_0}}
(\frac{l}{L_P})^2\frac{1}{\epsilon^2}=\delta
\end{eqnarray}
and we find:
\begin{eqnarray} \label{spectral} n &=&
1-\frac{1}{\pi y} (\frac{l}{L_p})^2 \frac{1}{\epsilon_i ^2}
\end{eqnarray}
The inflation would end ($\epsilon_f = 1$) when the ``slow roll''
parameter $\eta = \delta = 1$. We should evaluate the QCD coupling
constant $ \alpha_{S_0}(\mu) = \frac{4
\pi}{\beta_0\ln(\frac{\mu^2}{\Lambda_{QCD}^2})}$ at an energy
scale corresponding to the inflationary period. We take this to be
around the GUT scale $\sim 10^{15}GeV$ and
$\beta_0=11-\frac{2}{3}n_f=7$ whereas the weak logarithmic
dependence would assure the same order of magnitude for
$\alpha_{S_0}$ calculated at other larger scales. With
$\Lambda_{QCD}\sim 0.2 GeV$ \cite{data99} we estimate
$\alpha_{S_0}\sim 0.025$, and so we get
\begin{eqnarray} \label{loverL} (\frac{l}{L_P})^2&\sim& 10^{-1}
\end{eqnarray}
This is in disagreement with Bekenstein assumption that $L_P$ is
the shortest length scale in any physical theory. However, it
should be noted that Beckenstein's framework is very similar to
the dilatonic sector of string theory and it has been pointed out
in the context of string theories\cite{strings00} that there is no
need for a universal relation between the Planck and the string
scale. Furthermore, determining the order of magnitude of
$\frac{l}{L_P}$ is interesting in the context of these theories.

From (\ref{ei}), we have $\epsilon_i \sim 11$, and then using
(\ref{spectral}), we have $n=0.97$ which is within the range of
WMAP results \cite{wmap}.

Thus, we see that our model reproduces the results of the chaotic
inflationary scenario. However, the shape of the potential was not
put by hand, rather a gauge theory with a changing coupling
constant led naturally to it. Moreover, in typical chaotic models,
the inflaton field starts from very large values ($\phi_i \sim 15
M_{Pl}$) and ends at around $1 M_{Pl}$. One might suspect whether
field theory is reliable at such high energies. Nonetheless, this
problem is absent in our model since the large values have another
meaning in that they just refer to a reduction of the strong
coupling by around 10 times during the inflation.

Furthermore, chaotic inflations get a typical reheating of order
$T_{rh} \sim 10^{15} GeV$, and one might need to worry about the
relic problem. Similarly, equation (\ref{G2ei}) leads in our model
to a gluon condensate $|<G^2>|_i \sim 10^{62} GeV^4$ at the start
of inflation. From equation (\ref{epsevo}), we see that this
corresponds to an inflation time interval $\Delta t \sim
10^{-35}s$ satisfying the constraint (\ref{timecons}). If the
gluon condensate stays constant, as we assumed in our analysis, we
will have the same reheating temperature as in chaotic models
($T_{rh} \sim \rho(t_f)^{1/4}$). However, we should compare this
value for $<G^2>$ with its present value renormalized at GUT scale
$\sim 10^{15} GeV$ which can be calculated knowing its value at
$1GeV$ \cite{chamoun2000} and that the anomalous dimension of
$\alpha_S G^2$ is identical ly zero. We get
\begin{eqnarray}\label{condNow}<G^2(now, \mu\sim 10^{15} GeV)>\sim 1 GeV^4 \end{eqnarray} which
represents a decrease of $62$ orders of magnitude.

\begin{figure}[ht]
\begin{center}
\begin{tabular}{cc}
\epsfig{file=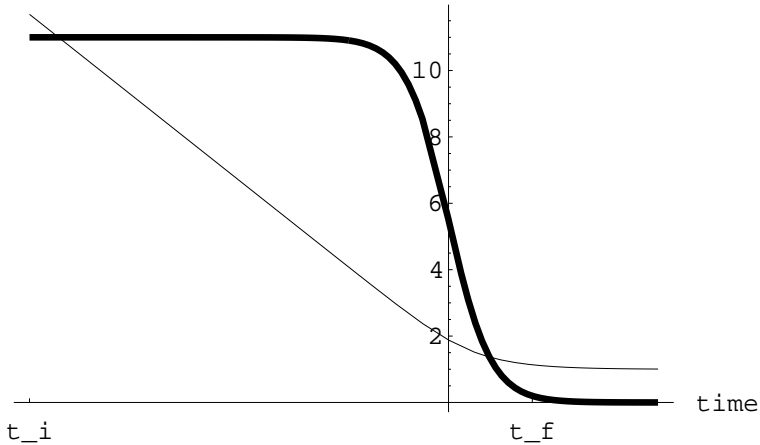}
\end{tabular}
\end{center}
\caption{Temporal evolution of the condensate $|<G^2>|$ (thick
line) and the $\epsilon$-field (thin line), for the choice $f(t) =
-\beta^2 \tanh ^2(\epsilon -1)$. The $<G^2>$ scale has been
adapted so that to visualize both graphs together.} \label{figure}
\end{figure}

This can give us the following possible picture for an exit
scenario. Lacking a clear theory for the non-perturbative
dynamical gluon condensate, we consider its value $|<G^2>|$
depending on energy, and thus implicitely on cosmological time, as
given by the standard RGE which turns it off logarithmically at
high energy. However, we can furthermore assume the condensate
value  to depend explicitely on time during inflation:
\[ <G^2(E,t)> = <G_0^2(E(t))> f(t)
\]
where $<G_0^2(E)>$ is the piece determined by the RGE. In order
that our model be consistent, the value of $<G^2>$ at the start of
inflation should be very huge and negative. The unknown function
$f(t)$ should be such that it varies slowly during most of the
inflationary era, to conform with an approximately constant value
of $<G^2>$, while at the end of inflation it causes a drastic drop
of the condensate value $<G^2>$ to around zero. The energy release
of this helps in reheating the universe, while reaching the value
$0$ leads to a minute ``trace-anomaly" energy mass density
(equation \ref{rhoT}) ending, thus, the inflation. The other types
of energy density would contribute to give the gluon condensate
its `small' positive value of (\ref{condNow}), and the subsequent
evolution is just the standard one given by RGE. Surely, this
phenomenological description needs to be tested and expanded into
a theory where the concept of symmetry breaking of such a phase
transition for the condensate $<G^2>$ provides the physical basis
for ending the inflation. Nonetheless, with a test function of the
form $f(t) = -\beta^2 \tanh ^2(\epsilon -1)$ with $\beta \sim
10^{31}$, one can integrate analytically the equation of motion,
and in ``slow roll'' regime we have $\epsilon = 1 + Arcsinh (\exp
[- \alpha \beta t])$ with $\alpha \beta \sim 10^{12} GeV$. The
graph in Fig. \ref{figure} shows the time evolutions of the
condensate and the $\epsilon$-field, which agree with the required
features. This example is meant to be just an existence example,
and the temporal dependence of the condensate could be of complete
different shape while the whole picture is still self-consistent.
The issue demands a detailed study for the condensate within an
underlying theory and we do not further it here. We hope this work
will stimulate interest in the subject.

\section*{Acknowledgements}
This work was supported in part by CONICET, Argentina. N. C.
recognizes economic support from TWAS.


\begin{thebibliography}{99}



\bibitem{guth81}A. H. Guth, {\em
Phys. Rev. } {\bf D 23}, 2, 347 (1981)

\bibitem{linde82}A. Linde, {\em
Phys. Lett. } {\bf B 108}, 389 (1982)

\bibitem{albrecht82} A. Albrecht and P. Steindhardt, {\em Phys. Rev.
Lett.} {\bf 48}, 1220 (1982)

\bibitem{linde83}A. Linde, {\em
Phys. Lett. } {\bf B 129}, 177 (1983)

\bibitem{linde94} A. Linde, D. Linde and A. Mezhlumian, {\em
Phys. Rev. } {\bf D 49}, 4, 1783 (1994)


\bibitem{deBernardis}
P. de Bernardis  et al., Nature {\bf 404}, 995 (2000).

\bibitem{MAXIMA}
S.Hanany et al.,Astrophys.J. 545 (2000) L5.

\bibitem{wmap}
H. V. Peiris {\it et al}, First Year Wilkinson Microwave
Anisotropy Probe (WMAP) Observations: Implications For Inflation,
ApJS, 148, (2003) 213, astro-ph/ 0302225

\bibitem{alexander01}
S. Alexander, R. Brandenberger and J. Magueijo, {\em Phys. Rev. }
{\bf D 67}, 081301 (2003), hep-th/0108190

\bibitem{khoury}
J. Khoury, B. Ovrut, N. Seiberg, P. Steindhardt and N. Turok, {\em
Phys. Rev. } {\bf D 65}, 086007  (2002)\\
J. Khoury, B. Ovrut, P. Steindhardt and N. Turok, {\em Phys. Rev.
} {\bf D 64}, 123522  (2001)

\bibitem{moffat93} J. Moffat, {\em
Int. J. P.} {\bf D 2}, 3, 351 (1993); Foundations of Physics, Vol.
23, 411 (1993)

\bibitem{albrecht99} A. Albrecht and J. Magueijo, {\em Phys. Rev.
} {\bf D 59}, 04, 3516 (1999)

\bibitem{barrow98}J. Barr
ow and J. Magueijo, {\em
Phys. Lett. } {\bf B 443}, 104 (1998)

\bibitem{chamoun2000} N. Chamoun, S. Landau and H. Vucetich,
{\em Phys. Lett.}{\bf B 504}, 1 (2001),
astro-ph/0008436

\bibitem{bekenstein82} J. D. Bekenstein, {\em Phys.
Rev. } {\bf D 25}, 6, 1527 (1982)

\bibitem{hawking00} S.W. Hawking, T. Hertog and H.S. Reall, {\em Phys.
Rev. } {\bf D 63}, 083504 (2001)
 hep- th/0010232,

\bibitem{itzykson80} C.I. Itzykson and J-B. Zuber {\em Quantum Field
Theory},
McGraw-Hill, 569 (1980)

\bibitem{greiner95} W. Greiner and A. Schafer, {\em Quantum Chromodynamics,
2nd ed},
 Springer, (1995)

\bibitem{guth01} A.H. Guth, astro-ph/0101507, {Proceedings of \em Cosmic
Questions 1999},
The New York Academy of Sciences Press

\bibitem{data99} {\em Particle Data Group}, {http://pdg.lbl.gov/}
(June 14, 2000)

\bibitem{barrow87} J. D. Barrow, {\em Phys. Rev.}
{\bf D 35}, 6, 1805 (1987)


\bibitem{hawking82} S. Hawking, {\em
Phys. Lett. } {\bf B 115}, 4, 295 (1982)

\bibitem{hawking85} S. Hawking, {\em
Phys. Lett. } {\bf B 150}, 5, 339 (1985)

\bibitem{brandenberger97} R. Brandenberger, {\em ICTP 1997  Summer
School in High Energy Physics and Cosmology}, ed. E. Gava {\em et
al}, World Scientific, (1997),\\
R. Brandenberger, {\em Invited lectures at TASI-94 Colorado}, (June 1994),
 astro-ph/9411049

\bibitem{mukhanov92} V. Mukhanov, H. Feldman and R. Brandenberger, {\em
Phys. Rep. },
{\bf 215}, 203 (1992)

\bibitem{strings00}C.P.Bachas, {\em
Class. Quant. Grav. } {\bf 17}, 951 (2000),
hep-th/0001093,\\
I.Antoniadis and B.Pioline, {\em Nucl. Phys. } {\bf B 550}, 41
(1999), hep-th/9902055



\end{thebibliography}
\end{document}